\documentclass{aa}

\usepackage{graphicx}
\usepackage{amsmath}

\input{tcilatex}

\begin{document}
\thesaurus{2(12.07.1; 08.02.3; 05.01.1)}

    \title{Trajectories of the images in binary microlensing}

\author{V.Bozza\thanks
{E-mail: valboz@sa.infn.it}}  \institute{Dipartimento di Scienze
Fisiche E.R. Caianiello,\\
 Universit\`{a} di Salerno, I-84081 Baronissi, Salerno, Italy.\\
 Istituto Nazionale di Fisica Nucleare, sezione di Napoli.}
\date{Received 3 july 2000/ Accepted 12 April 2001}
\maketitle

\begin{abstract}
We study in detail the trajectories followed by single images
during binary microlensing events. Starting from perturbative
resolutions of the lens equation, we explore the full parameter
space by continuity arguments. We see that the images created
during the caustic crossing can recombine with the others in
different ways. This leads to a new classification of the
microlensing events according to the behaviour of the images. We
show that the images involved in these combinations depend on the
folds that are crossed at the entry and at the exit from the
caustic. Non-trivial trajectories can be classified into four main
types. Some consequences for the motion of the center of light of
the source in astrometric measurements are also examined.
\keywords{Gravitational lensing - Binary stars - Astrometry}
\end{abstract}

\section{Introduction}

The theoretical study of binary lenses started in 1986, with the
paper by Schneider \& Wei{\ss}, where the equal masses case was
considered in detail. The caustics were derived and they started
to explore the non-trivial image dynamics underlying this model.
Their results were extended by Erdl \& Schneider (1993) and Witt
\& Petters (1993) to an arbitrary mass ratio. These works contain
two independent analytical derivations of the caustic structure of
the binary lens.

On the contrary, the direct inversion of the binary lens equation
can only be performed numerically, as it can be reduced to a fifth
order complex equation (Witt 1990). This point poses a major
difficulty in the extraction of analytical results about the
images of the binary lens. Two successful examples are the minimum
amplification during the caustic crossing, computed by Rhie (1997)
and the shape of the light curves in the caustic crossing by a
limb darkened source (Rhie \& Bennett 1999). In practice,
numerical methods are applied to fit experimental curves (Mao \&
di Stefano 1995; Albrow et al. 1999) and to make empirical
investigations about what should be expected in several cases (Di
Stefano \& Perna 1997; Dominik 1995, 1998). Obviously, this lack
of analytical results prevents a full understanding of the
problem, so that almost all we know of the images and their
magnification relies on numerical results whose generality cannot
be proved.

A new interest in the study of the trajectories of the images has
recently risen with the possibility of exploiting space
interferometry to measure the astrometric displacements of the
source stars being microlensed (Walker 1995; Dominik \& Sahu
2000). These researches are encouraged by several projects of
space missions that will bring the resolution down to few
microarcsecs (Allen, Peterson \& Shao 1997; Lindegren \& Perryman
1996). In the case of binary microlensing, astrometric
observations can lead to a complete break of the degenerations
affecting the light curves, since paths giving similar light
curves would be discriminated (Han, Chun \& Chang 1999; Gould \&
Han 2000). Following the motion of the center of light (CoL) of
the source can also help in the detection of extrasolar planets
with a more accurate determination of the parameters of the system
(Safizadeh, Dalal \& Griest 1999). It is likely that in the near
future it will be possible to follow the images in microlensing
events independently, at least when the source is not too large.

With this paper, we mean to bring new analytical knowledge to this
field, going beyond numerical studies, opening a discussion about
the generality of the results and studying the relations between
the observables and the parameters of the system. In Sect. 2, we
give analytical resolutions of the lens equation in special cases
where perturbative approximations are possible. In Sect. 3, we
review the gravitational lensing near cusps, looking at the local
structure of the lens equation. Then, in Sect. 4, we consider the
full lens equation and, by continuity arguments, we exploit the
results of Sect. 3 to investigate the trajectories of the images
globally. This leads to a new classification of the microlensing
events, according to the continuity of the trajectories of the
images involved in the caustic crossing. We shall also identify
the features distinguishing each topology from the others.
Finally, in Sect. 5, we give the summary.

\section{Perturbative resolutions of the binary lens equation}

Perturbative solutions show several interesting aspects of the
inversion problem, giving insight to some questions and opening
new ones. Moreover, they provide some starting points for a global
investigation of the images.

We introduce the Einstein radius of the total mass $M$:
\begin{equation}
R_{\mathrm{E}}=\sqrt{\frac{4GM }{c^{2}}
\frac{D_{\mathrm{LS}}D_{\mathrm{OL}}}{D_{\mathrm{OS}}}}.
\end{equation}
We indicate the coordinates in the lens plane normalized to
$R_\mathrm{E}$ by $\mathbf{x}=\left( x_1, x_2 \right)$ and the
coordinates in the source plane by $\mathbf{y}= \left( y_1, y_2
\right)$. All masses are measured in terms of $M$. We choose the
center of mass as the origin and put the two masses $m_1$, $m_2$
on the $x_1$ axis, separated by a distance $a$. We have
$m_1+m_2=1$ in our notations.

The lens equation reads
\begin{mathletters}
\begin{eqnarray}
&&y_1=x_1-\frac{m_1\left(x_1-am_2\right)}
{\left(x_1-am_2\right)^2+x_2^2} -\frac{m_2\left(x_1+am_1\right)}
{\left(x_1+am_1\right)^2+x_2^2} \\ &&y_2=x_2-\frac{m_1 x_2}
{\left(x_1-am_2\right)^2+x_2^2} -\frac{m_2x_2}
{\left(x_1+am_1\right)^2+x_2^2}.
\end{eqnarray} \label{Lens equation}
\end{mathletters}

\subsection{Source far from the lenses}

The simplest case, where perturbative resolution of the lens
equation is possible, is when the source is very far from the
lenses. We consider the inverse of the modulus of the source
position $|\mathbf{y}|$ as a perturbative parameter. In the limit
$|\mathbf{y}|$ going to infinity, the predominant
image is the source itself, i.e. $\mathbf{x}=\mathbf{y}$, since
the lens effects vanish. Expanding the lens equation (\ref{Lens
equation}) starting from this image, we get corrections to its
position in the form of a series in inverse powers of
$|\mathbf{y}|$. Stopping at the first order, we have
\begin{mathletters}
\begin{eqnarray}
&&x_1=y_1+\frac{y_1 }{y_1^2+y_2^2}\\ &&x_2=y_2+\frac{y_2
}{y_1^2+y_2^2}
\end{eqnarray}
\label{Far Principal}
\end{mathletters}
with positive parity.

We get the same solution if we expand the exact solutions for a
single lens with mass equal to the total mass of the system,
placed at the origin. In the following, this image will be
referred to as the principal image of the system.

Besides the principal image, two more expansions of the lens
equation are possible, starting from the two mass positions. They
lead to the same solutions we get by expanding the single lens
secondary images for each mass. We write here the secondary image
of the first mass. A similar expression holds for the other:
\begin{mathletters}
\begin{eqnarray}
&&x_1=am_2-\frac{y_1 m_1}{y_1^2+y_2^2}\\ &&x_2=-\frac{y_2
m_1}{y_1^2+y_2^2}.
\end{eqnarray}
\label{Far Secondary}
\end{mathletters}
They have negative parity.

So, at least when the source is far from the lenses, we can
identify a principal image and a secondary image for each mass.

\subsection{Planetary systems}

The images of planetary systems were already found in Bozza
(1999), where they were used to build analytical light curves for
planetary microlensing events. It is easy to build an analytic
expression for the center of light motion during a planetary
microlensing event as well.

The three images found in the limit $m_2 \ll m_1$ can be easily
reported to the far source labeling introduced in Sect. 2.1. Two
images are just the single lens images of the star, slightly
displaced by the planet. The third image is the secondary image of
the planet. The latter gives a negligible contribution to the
light curve and CoL trajectory.

\subsection{Wide binaries}

In this case, the perturbative parameter is the inverse of the
distance $a^{-1}$. We put the first lens in the origin and the
other in the position $\left( a, 0 \right)$.

For the principal image and the secondary image of the first mass,
we can start from the single lens images
\begin{equation}
\mathbf{I}^\pm=\frac{\mathbf{y}}{2}\left(1 \pm
\frac{\sqrt{4m_1+|\mathbf{y}|^2}}{|\mathbf{y}|} \right)
\label{Single lens images}
\end{equation}
and calculate the first order perturbations in the usual way. We
find
\begin{mathletters}
\begin{eqnarray}
&&\Delta I_1=\frac{m_2}{a\left(m_1^2-\left|\mathbf{I}
\right|^4\right)} \left[ m_1 \left( I_2^2- I_1^2 \right)+
\left|\mathbf{I} \right|^4
\right] \\ %
&& \Delta I_2=-\frac{2 m_1 m_2 I_1
I_2}{a\left(m_1^2-\left|\mathbf{I} \right|^4\right)}.
\end{eqnarray}
\end{mathletters}

For the secondary image of the second mass, we can proceed in the
same way as for the planetary image. Up to the second order, we
have
\begin{mathletters}
\begin{eqnarray}
&& I^{\mathrm{a}}_1=a+\frac{m_2}{a}+\frac{m_2}{a^2}y_1 \\ %
&& I^{\mathrm{a}}_2=-\frac{m_2}{a^2}y_2.
\end{eqnarray}
\end{mathletters}

Even if we can calculate positions and magnifications of all
images, the light curves and CoL trajectories built by these
quantities do not prove to be as good as those of the planetary
case. The problem is that to have any significant modulation on a
light curve or CoL trajectory, we need the source to pass very
close to the caustic of the first mass. But, in this regime, the
two main images lie very close to the critical curve of the first
mass where the magnification diverges. This fact rules out the
possibility of employing these results in a reliable way. In the
planetary case, on the contrary, when the source is close to the
planetary caustic, the two main images are far from the critical
curve and can be approximated efficiently.

Deriving the formulae for wide binaries, we have considered the
separation of the lenses to be much greater than all other
lengths, including $|y|$. Thus they are valid for a source close
to the first mass. If the source is far from the first mass as
well, then the formulae of Sect. 2.1 hold.

\subsection{Close binaries}

Finally, we consider the opposite situation, where the distance
between the two lenses is much smaller than all other distances.
When we let the separation $a$ go to zero, we obtain a single lens
with mass $1$ at the origin of our reference frame. So the
starting points for the calculations of the images are the two
images of this single lens
\begin{equation}
\mathbf{I}^\pm=\frac{\mathbf{y}}{2}\left(1 \pm
\frac{\sqrt{4+|\mathbf{y}|^2}}{|\mathbf{y}|} \right).
\end{equation}

Again, with the same procedure, we can find the perturbations to
the principal and secondary images. We define the quantities
\begin{mathletters}
\begin{eqnarray}
&& \Delta y_1=\frac{a^2 m_1 m_2 I_1\left(I_1^2-3I_2^2 \right)} {
|\mathbf{I}|^6}\\ %
&& \Delta y_2=\frac{a^2 m_1 m_2 I_2 \left(3I_1^2-I_2^2 \right)} {
|\mathbf{I}|^6}.
\end{eqnarray}
\end{mathletters}
Then, the perturbations to the images are given by
\begin{mathletters}
\begin{eqnarray}
& \Delta I_1^\pm=&\frac{\Delta y_1}{2} \left(1\pm
\frac{\sqrt{4m_1+|\mathbf{y}|^2}}{|\mathbf{y}|} \right) \nonumber
\\ && \mp \frac{2m_1 y_1 \left( y_1 \Delta y_1+y_2 \Delta y_2
\right)} {|\mathbf{y}|^3\sqrt{4m_1+|\mathbf{y}|^2}} \\ & \Delta
I_2^\pm=&\frac{\Delta y_2}{2} \left(1\pm
\frac{\sqrt{4m_1+|\mathbf{y}|^2}}{|\mathbf{y}|} \right) \nonumber
\\ && \mp \frac{2m_1 y_2 \left( y_1 \Delta y_1+y_2 \Delta y_2
\right)} {|\mathbf{y}|^3\sqrt{4m_1+|\mathbf{y}|^2}}
\end{eqnarray}
\label{Delta I}
\end{mathletters}

A third image of negative parity can be found if we start from the
origin. This central image has coordinates
\begin{mathletters}
\begin{eqnarray}
&& I^{\mathrm{c}}_1=a(m_2-m_1)+a^2 m_1 m_2y_1 \\ %
&& I^{\mathrm{c}}_2=-a^2 m_1 m_2y_2.
\end{eqnarray}
\end{mathletters}
but is highly demagnified.

Also for close binaries, apparent modulations to light curves and
CoL trajectories are achieved in a regime where the two main
images are too close to singularities. We cannot use these results
to study these curves analytically.

The secondary images of a close binary system are drastically
different from those of the far source regime. Rather than left
and right secondary images, we have a global secondary image and a
central one.

To understand the transition between the two regimes, consider two
lenses of equal mass. If the source is on the $y_2$ axis, i.e. on
the axis of symmetry of the system, the calculations of this
section show that the secondary and the central image would be
placed on the $x_2$ axis. If we move the source away along the
$y_2$ axis, in the limit of a far source, two secondary images are
formed close to the two lenses which are off-axis. Then, it seems
to be non-trivial how the transition between these two regimes
occurs. However, the secondary and the central image become very
close when $|y| \sim \frac{1}{a}$. This is the order of distance
of the secondary caustics in close binary systems (Bozza 2000a).
We will see in Sect. 4 that these small singularities play a
fundamental role in the transition between the two regimes.

\section{Lensing near cusps}

Having explored the lens equation in the cases where it is possible to resort
to analytical approximations, we want to bring our attention to the general
case. To get any comprehension of the global shape of the trajectories
of the images, it is necessary to look at the local behaviour first. As will
be explained in the next section, the cusps play an important role in this
comprehension. So, it is useful to review some general aspects of lensing
near cusps in order to develop further studies.

In the standard treatment (Schneider \& Wei{\ss} 1992), the cusp
is at the origin of the source plane, while the point of the
critical curve generating the cusp (cusp point) is at the origin
of the lens plane. With a suitable rotation, the lens map can be
generically written in the following form:
\begin{mathletters}
\begin{eqnarray}
&& y_1=cx_1+\frac{1}{2}bx_2^2 \\
&& y_2=bx_1x_2+ax_2^3,
\end{eqnarray}
\label{Cusp lens equation}
\end{mathletters}
where the coefficients $a$, $b$, $c$ are related to the
derivatives of the Fermat potential. The sign of the cusp, defined
as the sign of the quantity $2ac-b^2$ (Blandford \& Narayan 1986),
determines the parities of the images involved in the cusp
lensing.

The lens equation (\ref{Cusp lens equation}) can be exactly
solved. In particular, along the cusp axis (i.e. when $y_2=0$),
the solutions take a simple form:
\begin{mathletters}
\begin{eqnarray}
&& x_1=\frac{y_1}{c}, \; \; \;x_2=0 \\ %
&& x_1=\frac{2ay_1}{2ac-b^2}, \; x_2=+
\sqrt{\frac{2by_1}{b^2-2ac}}
\\ %
&& x_1=\frac{2ay_1}{2ac-b^2},  \; x_2=-
\sqrt{\frac{2by_1}{b^2-2ac}}
\end{eqnarray}
\end{mathletters}

\begin{figure}
 \resizebox{\hsize}{!}{\includegraphics{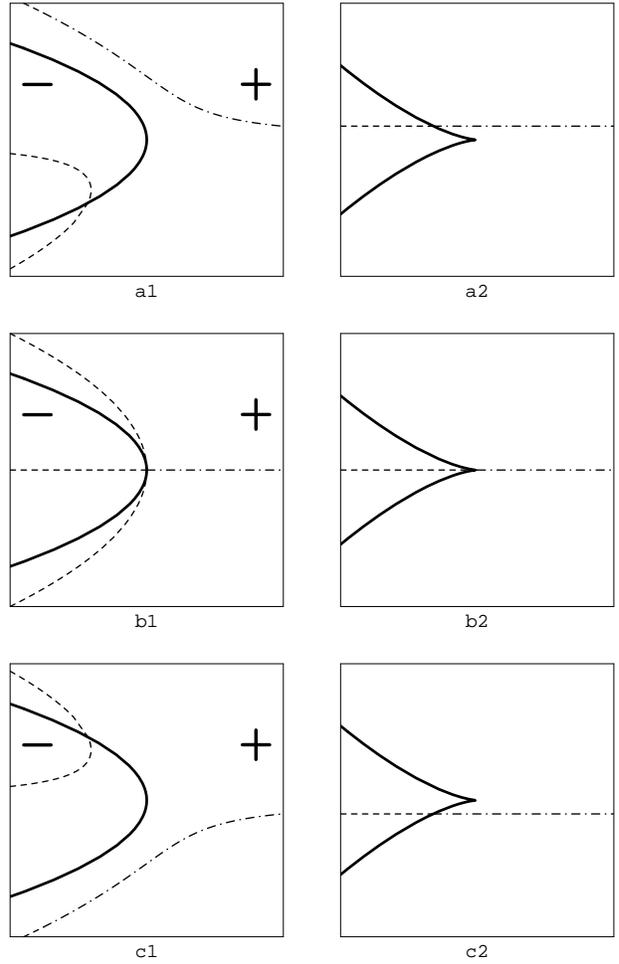}}
 \caption{Images in the neighborhood of a positive cusp. On the right,
 the cusp and the source trajectory are shown.
 On the left, the corresponding
 images and the critical curve are plotted.  The
 critical curve and the caustic are represented by the thick
 lines. The trajectories are described by dot-dashed lines
 before the caustic crossing and by dashed lines after. The sign of $\det J$ is also
 displayed. For a negative cusp, the sign of $\det J$ would be
 reversed and the images vertically reflected.
 (a1,a2): source trajectory crossing the fold above the cusp.
 (b1,b2):
 source trajectory crossing the cusp. (c1,c2): source trajectory crossing the
 fold below the cusp.}
 \label{Fig Positive cusp}
\end{figure}

We see that, for positive cusps, when $y_1>0$, there is one image
with positive parity on the $x_1$ axis. After the cusp
crossing (i.e. when $y_1<0$), this image has changed parity, while
two positive images are formed off-axis in specular
positions (Fig. \ref{Fig Positive cusp}b). The situation is
reversed for negative cusps.

If we slightly displace the source by a small constant $y_2$, the
caustic is crossed on a fold. Then, analyzing the exact solutions
and their parity, we see that the original image merges with the
top positive image, giving a fully continuous trajectory, without
changes of parity, moving along the critical curve on its positive
side on the top (Fig. \ref{Fig Positive cusp}a). The negative
parity branch of the former on-axis solution is slightly displaced
towards negative $x_2$. It starts below the cusp point on the
critical curve, together with the remaining positive image.

Of course, if we consider a small negative $y_2$, the plots are
reversed around the horizontal axis (Fig. \ref{Fig Positive
cusp}c). We see that the creation process occurs on the opposite
side of the fold crossing point (with respect to the cusp axis).
So, if we go from source trajectories crossing the caustic on the
upper fold to cusp crossing and then to trajectories crossing the
lower fold, we see that the continuous image changes side. It
passes on the top of the critical curve when the upper fold is
crossed and passes on the bottom when the lower fold is crossed.
Meanwhile, the creation point rises from below to above. A similar
process can be done for negative cusps.

\section{Global properties of the trajectories of the images in binary
microlensing}

After these preparatory sections, we are ready to compare with the
global properties of the image trajectories. It must be stressed
that we are considering microlensing of a point source and
consequently examine the trajectories of point images. This
discussion would apply to every area element of an extended image,
but not to the image as a whole, since different area elements may
follow quite different trajectories and it is impossible to define
the trajectory of the whole extended image.

To deduce global properties of point image trajectories from local
ones, we resort to several continuity arguments:

\begin{enumerate}
\item{
The lens application is everywhere continuous and locally
differentiable (except from the poles located on the two masses).
}
\item{
At non-critical points, it is locally invertible and its inverse
is locally continuous and differentiable. This means that, as long
as the source does not cross any caustics, we can follow each image
unambiguously. Images cannot jump or cross the critical curves.
The trajectories of the images change continuously with variations
in the source trajectory or in the parameters of the lenses,
without intersecting, as long as the source trajectory is not
self-intersecting (like in microlensing case where it is a
straight line). So, we can follow and identify each image until it
eventually reaches a critical point.
}
\item{
Let us turn to critical points. When the source reaches a fold,
two images are created or destroyed on the corresponding critical
point.  Continuous changes of the source trajectory imply
continuous changes of the critical point and continuous changes of
the image trajectories that meet there. Other image trajectories
cannot participate because at a fold, just two images meet, and a
transition process would need more than two images to participate
at the transition point. Therefore, such transitions can only
occur at higher singularities.
 }
\item{
When the crossing point moves from a fold to a cusp, then another
image participates in the destruction process. As illustrated in
Fig. \ref{Fig Positive cusp}, a cusp exchanges two images in the
creation/destruction on the folds arising from the cusp. The sign
of the images exchanged coincides with the sign of the cusp.  }
\item{
All arguments can be repeated if we consider continuous
variations in the parameters of the
lens equation. The images will be shifted continuously without
topological modifications of their trajectories if no higher order
singularities are met.
}
\end{enumerate}

From these properties, we obtain the following statement:

{\it The topology of the
trajectories of the images only depends on the folds
intercepted by the source trajectory.
The topology of the trajectories does not change when the parameters of the binary
lens are changed (if the folds crossed remain the same) even when
transition through different topologies of the critical curves
occur}.

This justifies a classification of the trajectories of the images
based on the folds crossed by the source at the entry and at the
exit from the caustic.

\subsection{Classification of image trajectories}

For the observations exposed in the previous subsection, we are
entitled to extend the labeling of Sect. 2 to each source
trajectory that can be continuously transported to infinity
without crossing any caustics. In fact, for these trajectories,
there is always one positive image to be identified with the
principal image. The two negative images start their trajectories
and end each at the same mass without exchanging, since this
exchange would need a discontinuous jump. So they can be
identified with the secondary images of each mass.

There are two kinds of source trajectories not crossing caustics
that are not homotopic to far source trajectories. In wide
binaries, if the source passes in the middle of the two caustics,
it cannot be transformed into a far source trajectory continuously
without intersecting any caustics. The two negative images live
inside the two separate critical curves and cannot mix. So, even
in this case, they can be identified with the secondary images of
the two masses. The other source trajectories non-homotopic to far
sources are those passing between the central caustic and one
secondary caustic of a close binary configuration. Since, in this
case, the two negative images live in the same critical curve, in
principle they can exchange. So we need further investigation
about what happens inside a secondary critical curve. We will come
back to this case in Sect. 4.1.4.

Once we have established that the discriminators in our
classification are the folds intercepted by the source, we can
choose the parameters of the binary system in the most convenient
way, since their specific value is not essential. The
natural choice is the equal masses case.

To investigate the shapes of the trajectories of the images in
caustic crossing events, we need some well known starting point
where a full analytical description is available. The full binary
lens equation can be exactly solved for a source on the axis
joining the two lenses, i.e. when $y_2=0$. In the equal masses
case, it can also be solved on the symmetry axis orthogonal to the
latter, i.e. when $y_1=0$. The fifth degree complex equation is
separated in a second degree and a third degree equation. The
solutions can be easily studied to find the trajectories of the
images in these particular cases. They will provide the basis for
the complete classification of the trajectories, so we discuss
them in detail.

\subsubsection{Source on the horizontal axis}

\begin{figure*}
 \resizebox{\hsize}{!}{\includegraphics{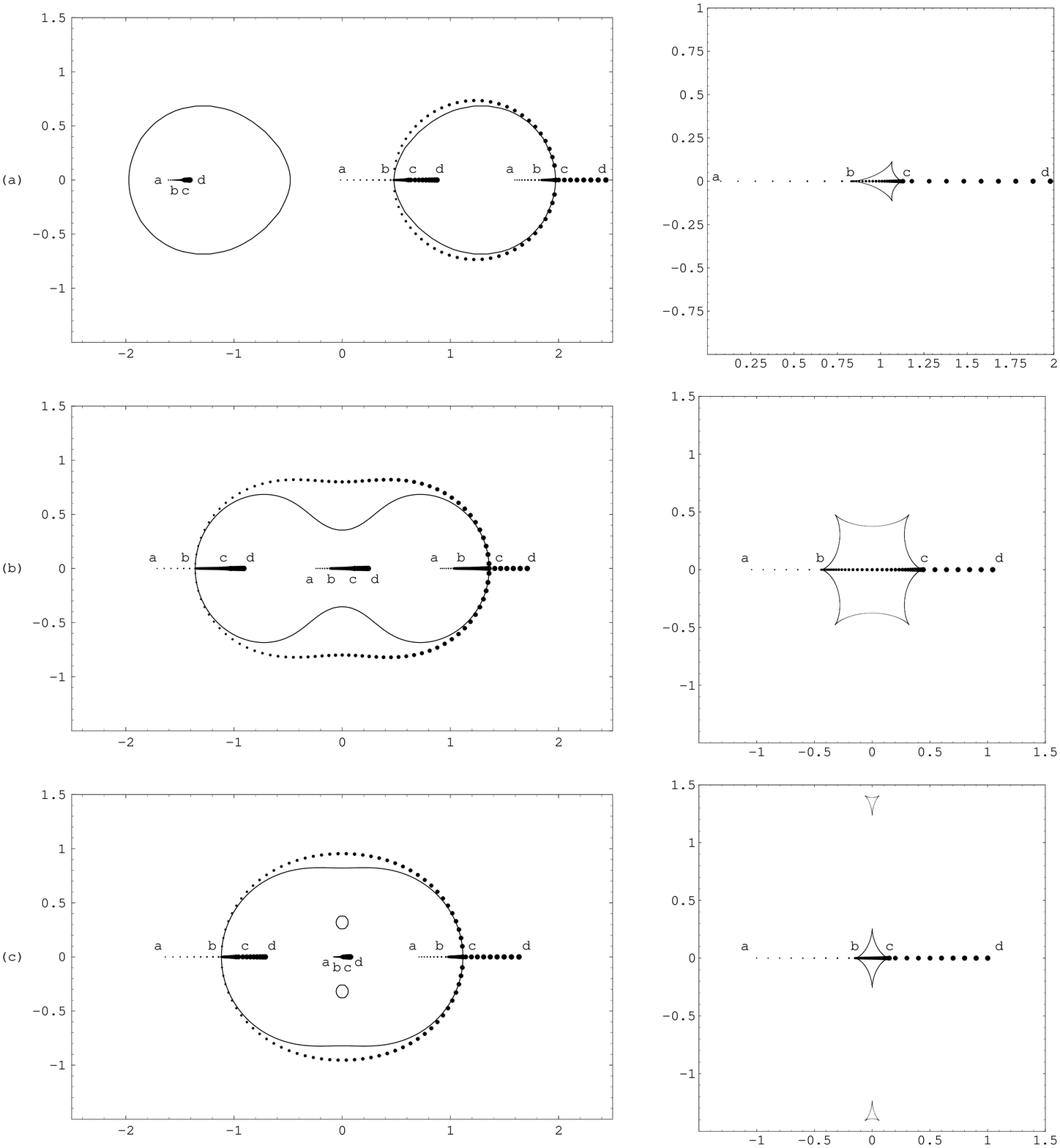}}
 \caption{Trajectories of the images (left) corresponding to the
 source trajectory on the right, lying on the horizontal axis.
 (a) wide binary, (b) intermediate binary, (c) close binary.
 To make the time sequence evident, we have
 displayed the positions of the images and the source by dots of
 increasing size. Moreover, letters mark the positions of the
 images and the source in the most important passages: $a$ is the
 starting point, $b$ is the entrance into the caustic, $c$ is the
 exit from the caustic, $d$ is the end of the trajectory.}
 \label{Fig Horizontal trajectories}
\end{figure*}

In Fig. \ref{Fig Horizontal trajectories} we have represented the
solutions for a source on the horizontal axis in the three
topologies of an equal masses binary system. We recall that the
cusps on the $y_1$ axis are positive and the others are negative,
as proved in (Bozza 2000b).

In the wide binary case (Fig. \ref{Fig Horizontal trajectories}a),
we have shown the right caustic crossing, since the other is
obviously symmetric. The source threads the caustic from left to
right. At the beginning, we have an image in the middle of the two
critical curves of positive parity and two negative parity images
in each of the two critical curves (time $a$). At time $b$ the
source meets the first cusp of the right caustic. It is a positive
cusp, so the positive image is substituted by a negative one,
inside the right critical curve and two positive images are formed
at the same point moving perpendicularly. At time $c$ the source
leaves the caustic. The two positive images meet again in the
right extremity of the critical curve to disappear. The original
negative image of the right critical curve is destroyed and a
positive image leaves the critical curve at the same point.
Finally (time $d$) there is a positive image on the right and
again a negative image in each critical curve.

\begin{figure*}
 \resizebox{\hsize}{!}{\includegraphics{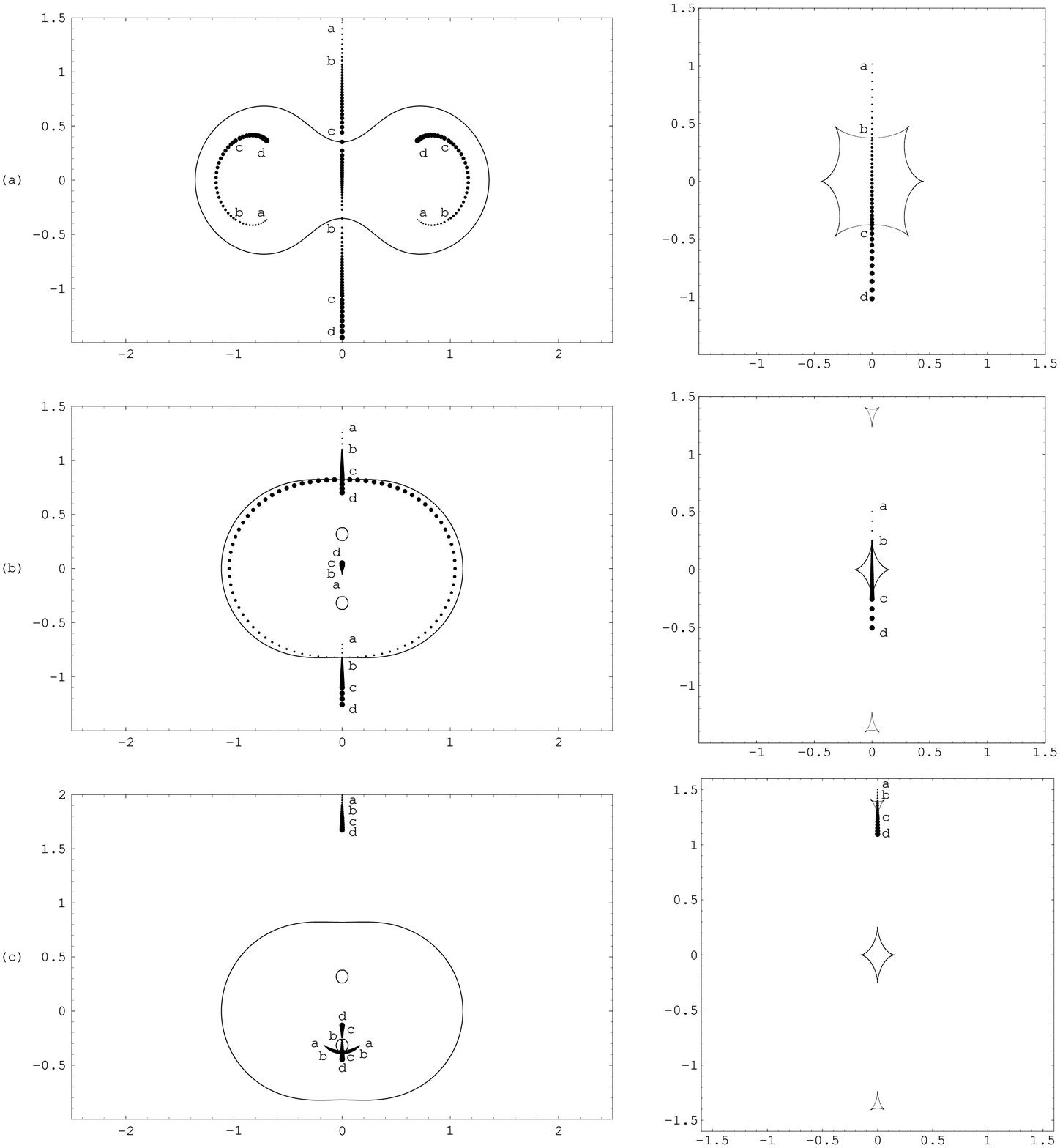}}
 \caption{Trajectories of the images for a source on the vertical
 axis: in the intermediate case (a),
 close binary case (central caustic crossing) (b),
 close binary case (secondary caustic crossing) (c).}
 \label{Fig Vertical trajectories}
\end{figure*}

In the same way, we can discuss the intermediate configuration
(Fig. \ref{Fig Horizontal trajectories}b). At the beginning (time
$a$), there is a positive image on the left and two negative
images close to each mass inside the critical curve. When the
source enters the caustic (time $b$), the positive image enters
the critical curve and two positive images are formed as in the
previous case. At time $c$, the two positive images, after having
passed the whole critical curve, one on the top and one on the
bottom, annihilate in the cusp point on the right. The right
negative image disappears and a positive one leaves the critical
curve. At the end (time $d$), we have the positive image on the
right and again two negative images close to each mass. We see
that the secondary image of the left mass has become the secondary
image of the right mass, without taking part in the
creation/destruction processes. So, in this case, there is an
exchange of secondary images between the two masses. One of the
two images is exchanged directly, while the other is exchanged via
creation/destruction processes.

For the close binary system (Fig. \ref{Fig Horizontal
trajectories}c), the discussion is similar to that of the
intermediate case, since the two cusps crossed are the same (in
the sense of continuous transformations of the lens configuration)
of the former case. So, here as well, there is an exchange of
secondary images.

\subsubsection{Sources on the vertical axis}

Now we come to the exact solutions on the vertical axis,
represented in Fig. \ref{Fig Vertical trajectories}. In the case
of wide binaries, it is not a caustic crossing curve, so we do not
take it into consideration.

We start from the intermediate case (Fig. \ref{Fig Vertical
trajectories}a). The source threads a fold in the entrance and in
the exit from the caustic. The two secondary images of the two
masses do not participate in the creation/destruction processes
and do not exchange. At time $b$, a pair of images with opposite
parities is created in the lower part of the critical curve. The
positive image just created survives at the end playing the role
of principal image when the source exits from the caustic (time
$c$). At this time, the former principal image is destroyed with
the negative image temporarily created. So, in this case, during
the caustic crossing time, it is not possible to identify a unique
principal image, because there are two positive images and each of
them becomes the principal image in one of the two asymptotic
regimes.

For close binary systems, the vertical trajectory meets both the
central and the two secondary caustics. The second plot (Fig.
\ref{Fig Vertical trajectories}b) deals with the central caustic
crossing. At the beginning (time $a$), there is a positive image,
a negative image at the centre to be identified with the central
image of Sect. 2.4 and a negative image close to the opposite
boundary of the main critical curve (global secondary image of
Sect. 2.4). At time $b$, the source meets a negative cusp. The
secondary image leaves the place to a positive image, and two
negative images are formed. These negative images travel to the
top, where they annihilate at time $c$. At the same point, the
former principal image is destroyed and a negative one is born,
playing the role of
global secondary image at the end. The central image is not
involved in any process.

Finally, we deal with the secondary caustic crossing (Fig.
\ref{Fig Vertical trajectories}c). At time $a$, the source is
above the top secondary caustic. There is a principal image and a
secondary image for each mass. These can be unambiguously
associated with their masses, since, for $y_2$ going to infinity,
they reach them. At time $b$ the source reaches the secondary
caustic in the upper fold. A pair of images is created at the top
of the lower oval. The negative image just created goes towards
the centre and finally plays the role of central image. The
positive one goes to the bottom of the oval and at time $c$, when
the source exits the secondary caustic through the lower negative
cusp, it becomes negative, finally playing the role of global
secondary image. The two former partial secondary images are
destroyed at the cusp point at time $c$. So, it is clear now how
the transition between the far source regime and the close binary
regime happens, at least in the vertical trajectory case: first,
the central image is created and then its temporary positive
partner converts to the global secondary image while the partial
secondary images disappear.

\subsubsection{Fold crossing trajectories}

We have previously stated that the topology of the paths of the
images depends on the folds crossed by the source during the
microlensing event. We will derive these topologies starting from
those of the exact solutions on the two axes of the system, by
slightly perturbing the trajectory and applying the discussion of
Sect. 3 to resolve the lens equation in the neighborhood of the
cusps.

In general, a source trajectory can intersect the caustics more
than once. In each caustic crossing, it has to enter through a
fold and exit through another fold (it cannot be the same fold,
since the folds seen from the inside are convex). So, we can
decompose a trajectory in a succession of simple caustic crossings
and study them separately. For this reason, we shall examine, in
what follows, just simple caustic crossings. Multiple caustic
crossings are obtained by putting together two or more simple
crossings.

The two caustics of the wide binaries are equivalent for
reflection on the vertical axis. Let's consider the right one.
Each caustic has four folds. We number them from the upper left to
the lower left clockwise. Obviously, we can pair the folds in six
ways. Two pairs (3-4, 2-4) can be obtained from other two pairs
(1-2, 1-3) by reflection around the horizontal axis, so we have to
consider four possibilities for fold crossings. The trajectories
intersecting the caustic in these pairs are shown in Fig. \ref{Fig
Wide binary}.

\begin{figure*}
 \resizebox{\hsize}{!}{\includegraphics{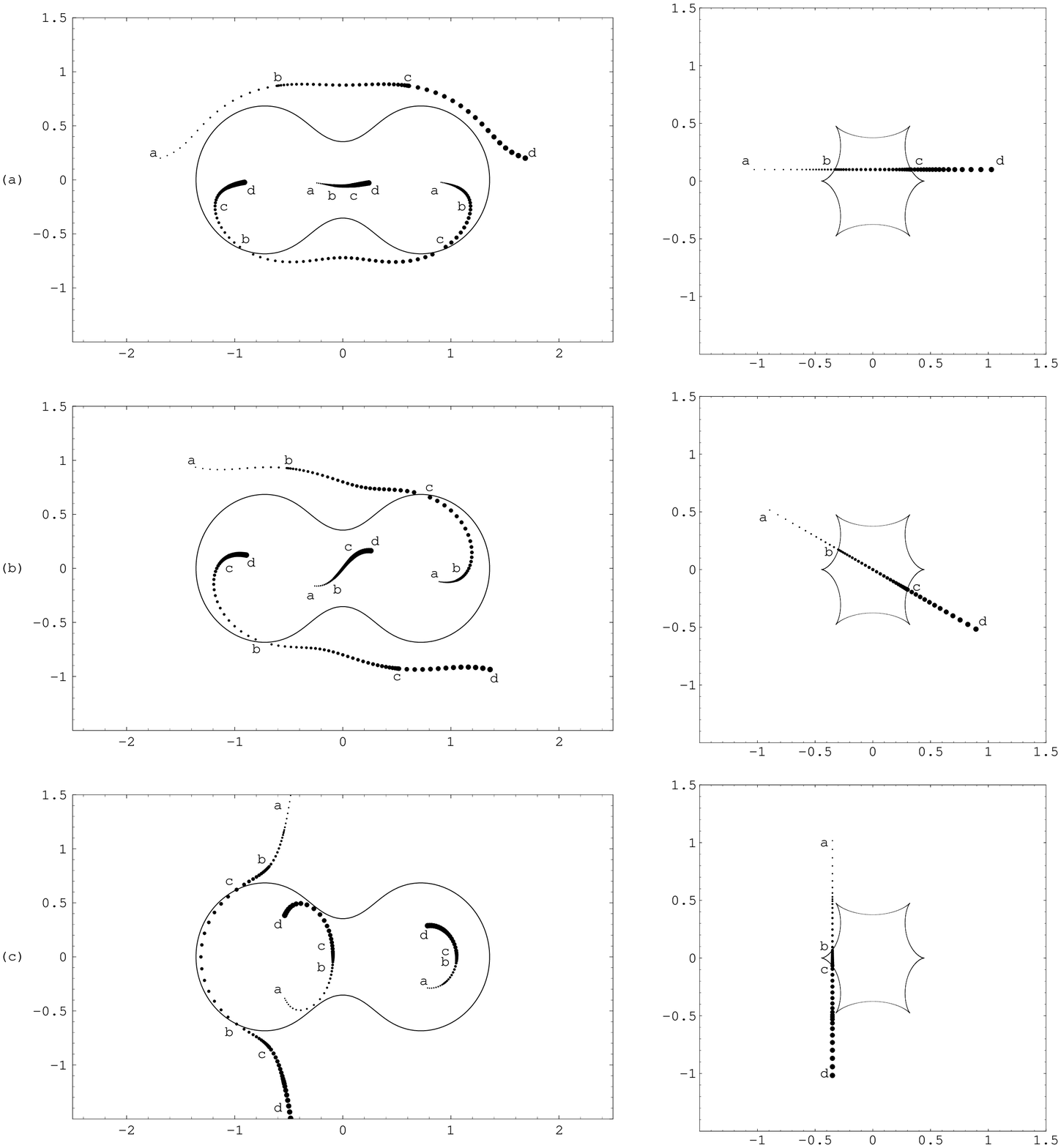}}
 \caption{Three of the six inequivalent fold crossings for the caustic of
  intermediate binaries. (a) 1-3, (b) 1-4, (c) 1-6.}
 \label{Fig Intermediate binary 1}
\end{figure*}

\begin{figure*}
 \resizebox{\hsize}{!}{\includegraphics{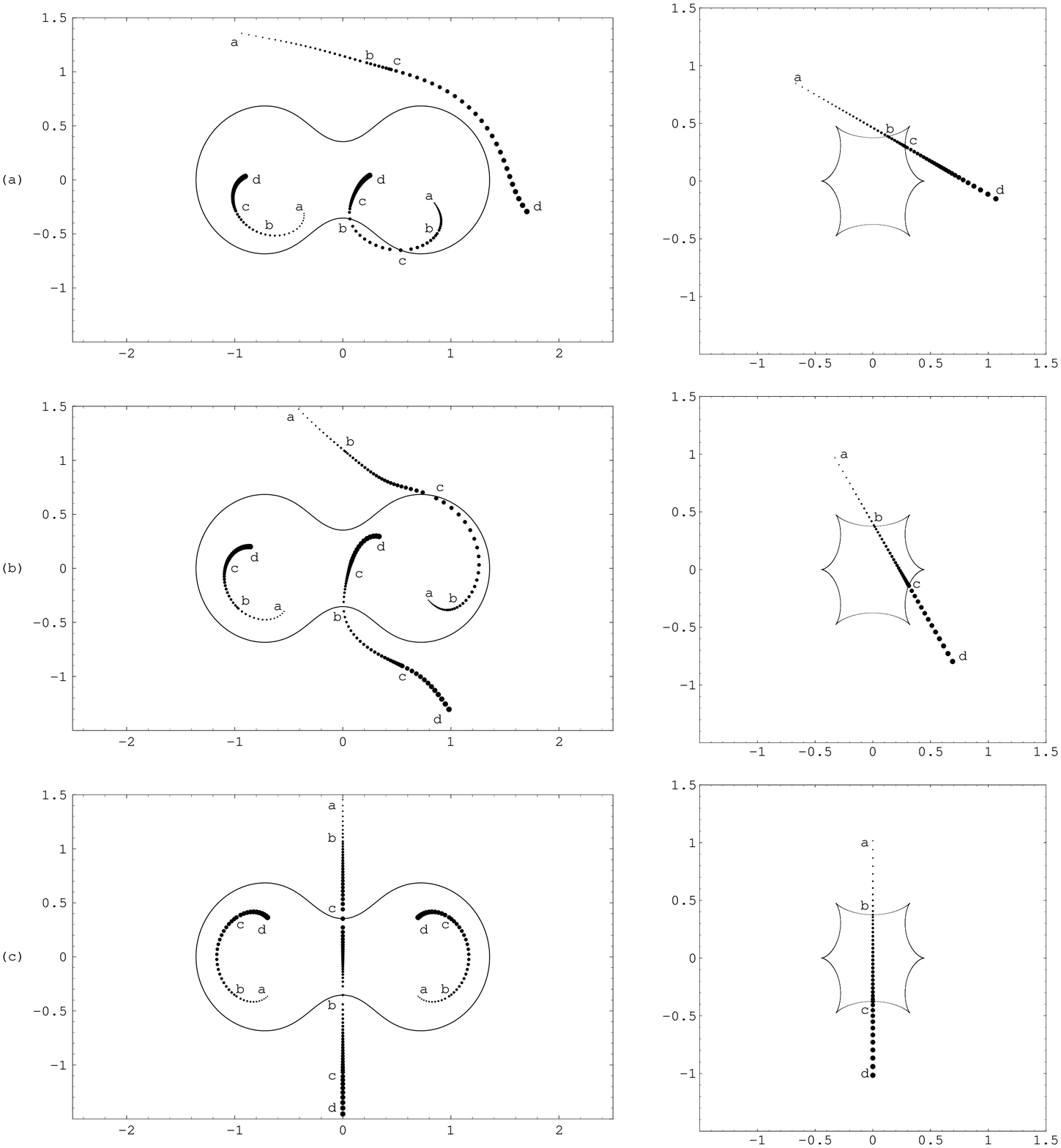}}
 \caption{The other three of the six inequivalent fold crossings for the caustic of
  intermediate binaries. (a) 2-3, (b) 2-4, (c) 2-5.}
 \label{Fig Intermediate binary 2}
\end{figure*}

The caustic of the intermediate configuration has six folds. We
number them in the same way as the wide binary caustic. By
considering equivalent pairs that can be obtained by reflection on
either axis, out of the fifteen pairs, we are left with six pairs
(1-3, 1-4, 1-6, 2-3, 2-4, 2-5). The first three trajectories are
shown in Fig. \ref{Fig Intermediate binary 1} and the other three
in Fig. \ref{Fig Intermediate binary 2}.

For the central caustic of close binaries there are three
possibilities (1-2, 1-3, 1-4), shown in Fig. \ref{Fig Close
central binary}. We number the folds of the upper secondary
caustic starting from the upper, clockwise. So we have two other
possible fold crossings (5-6, 6-7) (Fig. \ref{Fig Close secondary
binary}).

\begin{figure*}
 \resizebox{\hsize}{!}{\includegraphics{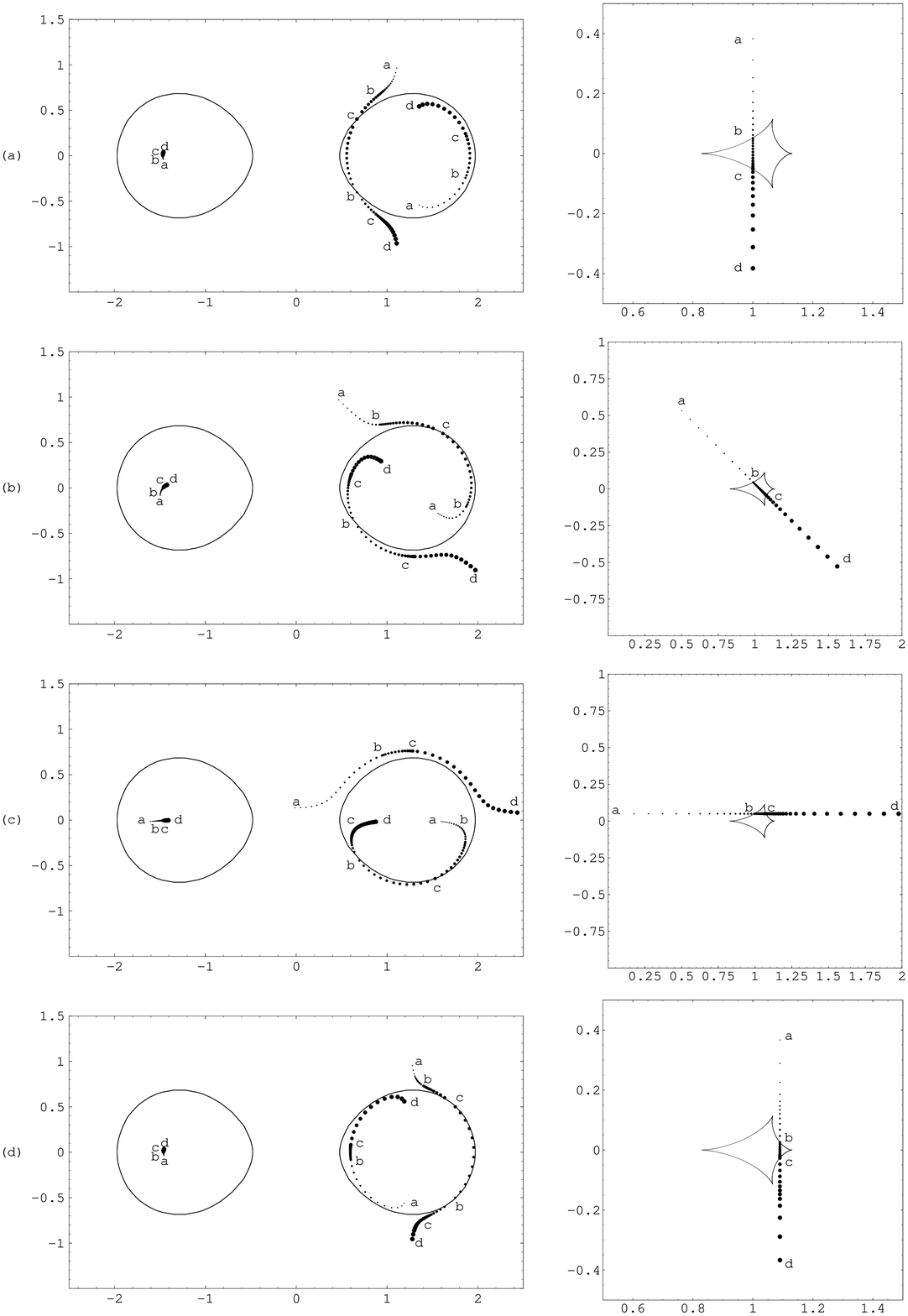}}
 \caption{The four inequivalent fold crossings in wide binaries. (a) 1-4,
 (b) 1-3, (c) 1-2, (d) 2-4.}
 \label{Fig Wide binary}
\end{figure*}

\begin{figure*}
 \resizebox{\hsize}{!}{\includegraphics{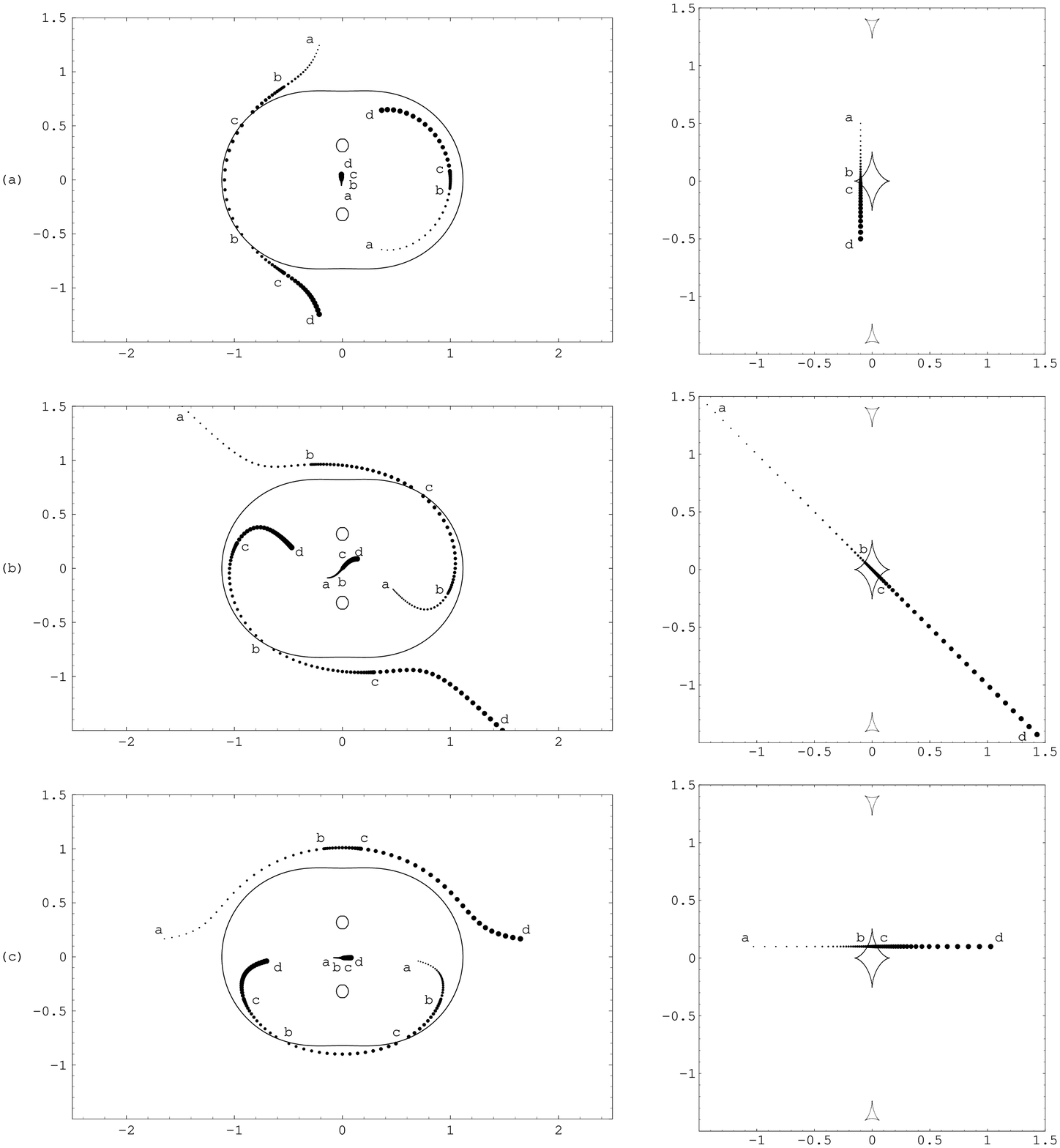}}
 \caption{The three inequivalent fold crossings for the central caustic of
  close binaries. (a) 1-4, (b) 1-3, (c) 1-2.}
 \label{Fig Close central binary}
\end{figure*}

\begin{figure*}
 \resizebox{\hsize}{!}{\includegraphics{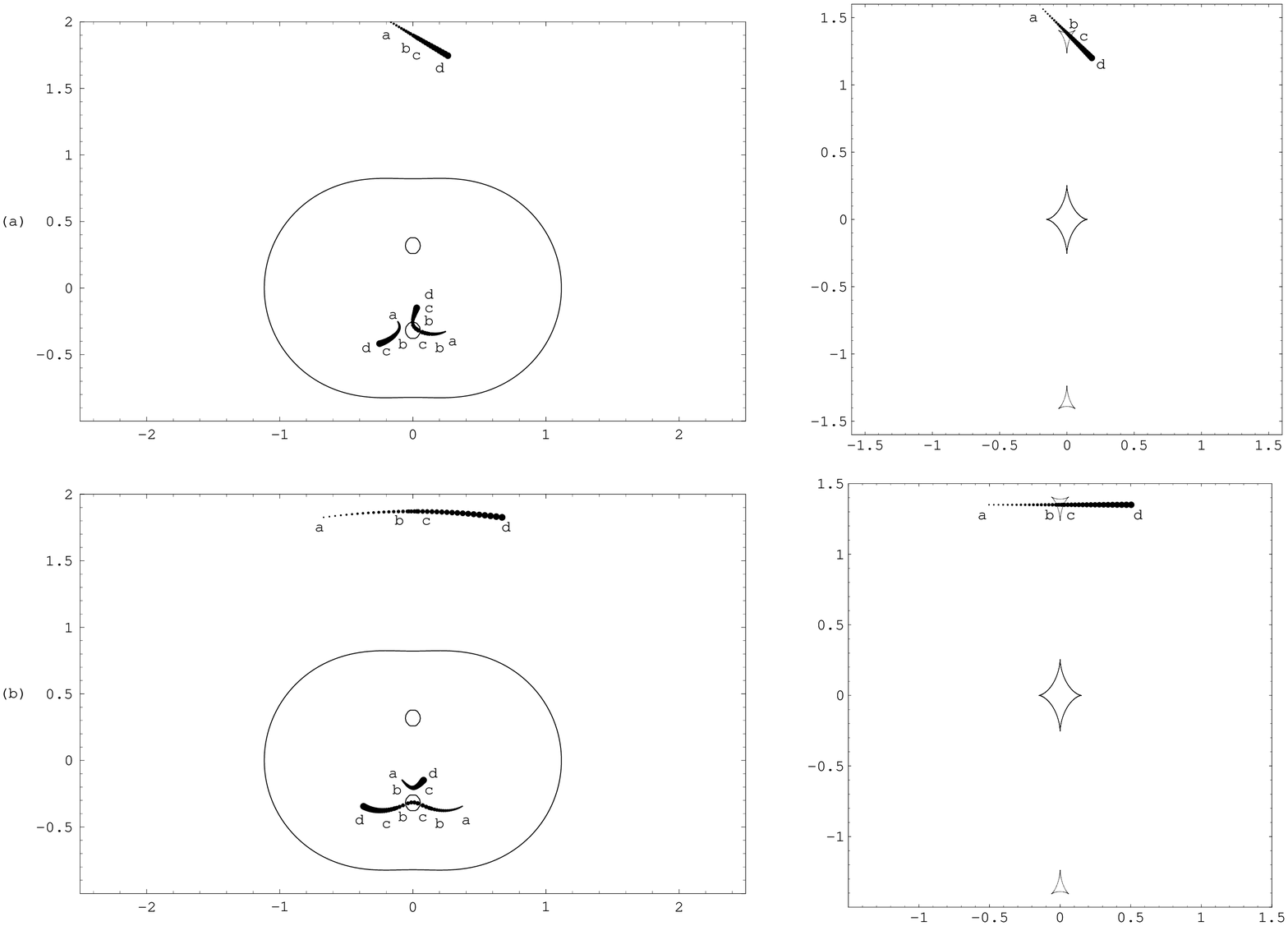}}
 \caption{The two inequivalent fold crossings for the upper secondary caustic of
  close binaries. (a) 5-6, (b) 6-7.}
 \label{Fig Close secondary binary}
\end{figure*}

Exploiting all the considerations done up to now, we are able to
discuss these trajectories and derive their properties
analytically.

The first fold crossing we want to discuss is wide binary 1-2,
represented in Fig. \ref{Fig Wide binary}c. It can be obtained by
slightly displacing the horizontal trajectory (shown in Fig.
\ref{Fig Horizontal trajectories}a) towards the top. As the two
cusps crossed in the horizontal trajectory are positive, following
the discussion of Sect. 3 about the behaviour of positive cusps,
we can say that the upper positive image created in the caustic
crossing has to join with the positive image living before and
after the crossing to form a unique continuous image. At the first
fold crossing (time $b$), the two opposite images have to be
created in the $y_2<0$ half of the critical curve. At the second
crossing (time $c$), the extra positive image is destroyed in the
same half of the critical curve, near the right cusp point. We can
observe that in this kind of fold crossing the principal image can
be continuously followed from the beginning to the end of the
event, even during the caustic crossing. The right secondary image
passes from right to left through a creation/destruction process
with an extra positive image. The left secondary image is
unaffected.

Modifying the parameters of the system, we can derive other
significant fold crossings from the wide binary 1-2. The
intermediate binary 2-3 is obtained by lowering the separation
between the lenses. The upper fold of the intermediate caustic is
the result of the merging of wide binary fold 1 of the right
caustic with the homologous of the left caustic. So, the
trajectories of the images will not be topologically changed
during the transition between the two regimes. So, in the
intermediate binary 2-3 crossing, represented in Fig. \ref{Fig
Intermediate binary 2}a, there is a continuous principal image, a
left secondary image unaffected and a right secondary image
jumping via creation of a positive image.

The wide binary 1-3 (Fig. \ref{Fig Wide binary}b) is derived by
displacing the horizontal trajectory by a $y_2=\epsilon (b-y_1)$
with $\epsilon$ small and $b$ is the position of the right
caustic. In this way, the left positive cusp is skipped from above
and the right from below. According to Sect. 3, the upper positive
image appearing during caustic crossing has to merge with the left
positive image existing before the crossing. This image has to
disappear at time $c$, where it meets the former right secondary
image. At the same time, the lower positive image appearing during
the caustic crossing merges with the right positive image existing
after the crossing. This image is created at time $b$ with the
final right secondary image. So, in this case, the situation is
more complicated, since the original principal image is destroyed
and another positive image takes its place. The same happens for
the right secondary image. So, we can say that at time $b$ a new
pair of principal and right secondary images are formed and the
preexisting ones are destroyed at time $c$.

From wide binary 1-3, we can derive the intermediate binary 2-4
(Fig. \ref{Fig Intermediate binary 2}b) in the same way as before
for the intermediate 2-3.

The intermediate binary 1-3 is derived from the horizontal
trajectory (Fig. \ref{Fig Horizontal trajectories}b) adding a
positive $y_2$. Then, as in wide binary 1-2, the positive
extra-image appearing during the caustic crossing joins with the
positive images existing before and after to give a fully
continuous principal image. The left secondary image becomes the
secondary image of the right mass, as in the horizontal case,
since it is not involved in singularities. At time $b$, a pair of
opposite images is formed near the left cusp point. The negative
image just produced becomes the left secondary image, while the
positive annihilates with the former right secondary image at time
$c$ near the right cusp point.

The close binary 1-2 crossing (Fig. \ref{Fig Close central
binary}c) involves the same folds of intermediate binary 1-3,
since the transition to close binaries separates the folds 1 and 3
of the intermediate caustic in two, with the separation of the
secondary caustics from the central one. So, the discussion is
similar to the previous. There is a continuous principal image.
The left secondary image becomes the right one, but we can also
identify it with the central image of this phase. The global
secondary image is broken in two pieces joined by a temporary
positive image.

The intermediate binary 1-4 is derived by the horizontal
trajectory by taking $y_2=-\epsilon y_1$ with $\epsilon$ small
(Fig. \ref{Fig Intermediate binary 1}b). The upper positive image
during the caustic crossing merges with the positive one before
the crossing. So, the principal image is destroyed at time $c$
with the right secondary image. The lower positive image joins
with the one after the crossing. So the late time principal image
is born at time $b$ along with the final left secondary image. The
initial left secondary image becomes the right one.

The discussion is the same for close binary 1-3, that can be
obtained from the latter case (Fig. \ref{Fig Close central
binary}b).

The intermediate 2-5 fold crossing (Fig. \ref{Fig Intermediate
binary 2}c) is described analytically by the vertical trajectory
(Fig. \ref{Fig Vertical trajectories}a), already discussed.

The wide binary 1-4 case, shown in Fig. \ref{Fig Wide binary}a,
involves the same folds of the intermediate 2-5 case after their
breaking in two pieces. So, in this case as well, the two
secondary images are unaffected, while the principal image is
broken into two pieces joined by a temporary negative image.

To attain the close binary 1-4 fold crossing, we have to displace
the vertical trajectory of Fig. \ref{Fig Vertical trajectories}b
towards the left. The two cusps intercepted were negative, so the
right negative image living during the caustic crossing joins with
the two pieces of the global secondary image to give a fully
continuous trajectory. The central image remains unaffected. The
left negative image deals with the joining of the two pieces of
the principal image which remains interrupted. So, at time $b$,
two opposite images are created at the bottom, on the left of the
cusp point. The positive image becomes the final principal image,
while the negative one travels to the top to destroy the former
principal image at time $c$.

From this case, we can obtain, by continuous transformation of the
separation between the lenses, the intermediate binary 1-6 (Fig.
\ref{Fig Intermediate binary 1}c). The roles of the two unaffected
secondary images is here more defined. We have to remember that
for a full definition of the two negative images in close binaries
we have to consider the secondary caustic crossing.

The wide binary 2-3 case (Fig. \ref{Fig Wide binary}d) is a
transformation of the close binary 2-3 crossing, which is the
reflection of 1-4 crossing, examined before. So, also in this
case, the two negative images are not involved and the principal
image is broken.

\begin{table*}
\begin{tabular}{|l|cccc|}
   \hline
    \bf{Source trajectory} & \bf{Breaking} & \bf{Exchange} & \bf{Ext. exchange} & \bf{Pair subst.} \\
    \hline
  \bf{Non caustic crossing} &  &  &  &  \\
  \hspace{0.5cm} Homotopic to far source &  &  &  &  \\
  \hspace{0.5cm} Non homotopic in wide binaries &  &  &  &  \\
   \hspace{0.5cm} Non homotopic in close binaries &  & L$\rightarrow$R, R$\rightarrow$L &  &  \\
  \bf{Caustic crossing} &  &  &  &  \\
  \hspace{0.25cm} \it{Wide binaries} &  &  &  &  \\
  \hspace{0.5cm} Folds 1-2 & R &  &  &  \\
  \hspace{0.5cm} Folds 1-3 &  &  &  & PR \\
  \hspace{0.5cm} Folds 1-4 & P &  &  &  \\
  \hspace{0.5cm} Folds 2-3 & P &  &  &  \\
  \hspace{0.25cm} \it{Intermediate binaries} &  &  &  &  \\
  \hspace{0.5cm} Folds 1-3 &  & L$\rightarrow$R & R$\rightarrow$L &  \\
  \hspace{0.5cm} Folds 1-4 &  & L$\rightarrow$R &  & PR$\rightarrow$PL \\
  \hspace{0.5cm} Folds 1-6 & P &  &  &  \\
  \hspace{0.5cm} Folds 2-3 & R &  &  &  \\
  \hspace{0.5cm} Folds 2-4 &  &  &  & PR \\
  \hspace{0.5cm} Folds 2-5 & P &  &  &  \\
  \hspace{0.25cm} \it{Close binaries} &  &  &  &  \\
  \hspace{0.5cm} Folds 1-2 &  & L$\rightarrow$R & R$\rightarrow$L &  \\
  \hspace{0.5cm} Folds 1-3 &  & L$\rightarrow$R &  & PR$\rightarrow$PL \\
  \hspace{0.5cm} Folds 1-4 & P &  &  &  \\
  \hspace{0.5cm} Folds 5-6 & R &  &  &  \\
  \hspace{0.5cm} Folds 6-7 &  & L$\rightarrow$R & R$\rightarrow$L & \\
  \hline
\end{tabular}
  \caption{Summary of the classification of the trajectories of the images.}
\end{table*}

Last, we consider the two kinds of secondary caustic crossings.
The 5-6 crossing (Fig. \ref{Fig Close secondary binary}a) can be
derived from the vertical trajectory (Fig. \ref{Fig Vertical
trajectories}c) adding a small positive $y_1$. The only cusp to
resolve is negative. At the beginning, we have a secondary image
for each mass (time $a$). At time $b$, the fold 5 is crossed and
two images are formed on the top of the lower oval, as in the
vertical trajectory. The negative image just formed becomes the
central image, while the positive image annihilates with the right
secondary image, according to the negative cusp resolution, as
discussed in Sect.3. The initial left secondary image is no longer
involved in the negative cusp destruction and becomes the global
secondary image. Finally, the transition between far source and
close binary regime is achieved.

The secondary caustic crossing 6-7 (Fig. \ref{Fig Close secondary
binary}b) can be considered as a transformation from the
intermediate 1-3 case, since the folds involved are pieces of the
folds 1 and 3 broken after the transition to the close binary
configuration. The principal image and the central image are
unaffected and the global secondary image is broken in two pieces
joined by a positive temporary image residing in the small lower
oval.

\subsubsection{Outlook and general considerations}

After the study of all possible fold crossing curves, we can
complete our classification by considering the non caustic
crossing source trajectory passing between the central and the
secondary caustic of the close binary system. If we start from the
6-7 crossing (Fig. \ref{Fig Close secondary binary}b), we can
obtain our trajectory by pushing the trajectory below the
secondary caustic. The source will no longer intersect the caustic
and the global secondary image will become continuous. So, the
difference between this trajectory and a non caustic crossing one
homotopic to far source trajectories is in the exchange between
the two partial secondary images. This is the only exchange
between secondary images where none of the two images is involved
in creation/destruction processes.

Looking at Figs. \ref{Fig Wide binary}-\ref{Fig Close secondary
binary}, there exist four possible events characterizing a
non-trivial trajectory.

\begin{enumerate}
\item{
Breaking: The image path is separated into two pieces joined by a
temporary image during the caustic crossing. }
\item{
Exchange: Continuous passage of a secondary image from one mass to
the other. }
\item{
External exchange: Exchange broken by the presence of a temporary
image(combination of Breaking and Exchange).}
\item{
Pair substitution: Two images of opposite parities are destroyed
at time $c$ while a new pair, substituting the former ones, is
created at time $b$.}
\end{enumerate}

Table 1 sums up the outcomes of our classification for all
possible trajectories, according to the scheme here explained.
$P$, $L$, $R$ stand for principal, left and right secondary image
respectively.

About the continuity of the principal image, it is possible to
draw a general conclusion. In fact, if a positive image has to be
principal in both asymptotic regimes, it has to thread the
horizontal axis (apart from the particular case of a source
trajectory parallel to this axis). Obviously, the candidate to the
principal image has to thread the horizontal axis out of the
critical curves. But, as we have seen discussing the source
trajectory along the $y_1$ axis, the image positions on the $x_1$
axis outside of the critical curve all correspond to sources on
the $y_1$ axis outside the caustics. So, we can say that all
source trajectories threading the $y_1$ axis inside the caustics
possess a discontinuous principal image. The converse is also true
if we look at the outcomes of our classification.

As regards the exchange of secondary images, we can see that it
happens when the two folds intercepted lie one on the left and the
other on the right of the caustic in intermediate and close
binaries. If one of the intercepted folds is the top or the bottom
one, no exchange occurs. The non-crossing source trajectory
passing between a secondary and the central caustic in close
binaries also manifests an exchange between the two secondary
images.

In no case are two images created at time $b$ destroyed together
at time $c$. The pairs of destroyed images are never the two
created previously. So, if we look at the union of all image
trajectories, forgetting about time ordering, we will see three
continuous curved lines. The extremities of these three curves are
those of the far source regime: no other possible start or end
points are available. Normally, the three curves will start and
end at the same pole (we can treat infinity as an effective pole
of the map in a one-point compactification of the image plane on a
sphere), resulting in three loops. If an exchange or a pair
substitution occurs, one curve will start on a pole and end on
another, while the curve starting from the latter pole will end on
the first. In this case, therefore, there will be two loops, since
two curves will be joined together through the poles. Finally, if
a pair substitution and an exchange occur simultaneously, there
will be just one loop, starting from infinity, passing through the
two masses and going back to infinity.

\section{Summary}

The behaviour of the images of gravitational lenses is almost an
untouched field of investigation. The difficulties in the
inversion of the lens equation is an obstacle to analytic
statements. In this paper, we have studied the images produced by
a binary lens. We started from particular cases where analytic
solutions were available; then, studying the lens equation near
cusps, we extended our results by continuity. In this way, we have
attained a full investigation of images in microlensing events,
with a complete classification of the topologies of the
trajectories and an identification of the four phenomena related
to topology changes. Many of our results have a more general
validity and could be extended to other lenses just by examining
the far source regime and the caustic structure which is, in
general, easier to derive than the images. So, our method gives
new insights into the study of the lens equation and the
topologies of the trajectories of the images.

It would also be very interesting to try to exploit these results
in the study of CoL motion in order to have an immediate
connection to a quantity that will become observable in the near
future. In principle, a complete classification of the light
curves and the CoL motion with respect to the folds intercepted by
the source is possible and would be a very useful tool for the
interpretation of a binary microlensing event. The classification
presented in this work is an important first step towards these
objectives.

\begin{acknowledgements}
I would like to thank Gaetano Scarpetta and Salvatore Capozziello
and Martin Dominik for their comments on the manuscript.

Work supported by the European Social Fund through PO by MURST.
\end{acknowledgements}

\end{document}